\documentclass[5p,twocolumn]{elsarticle}
\usepackage{graphicx,latexsym}
\usepackage{dcolumn}
\usepackage{amssymb,amsmath,bm}
\usepackage{subfigure}
\usepackage{braket}
\usepackage{siunitx}
\usepackage{array}
\usepackage{physics}
\usepackage{float}
\usepackage[nopar]{lipsum}
\usepackage{caption}
\usepackage{multirow}
\usepackage{bigstrut}
\usepackage[super]{nth}

\usepackage{hyperref}
\hypersetup{
    pdfnewwindow=true,       % links in new window
    colorlinks=true,         % false: boxed links; true: colored links
    linkcolor=blue,          % color of internal links
    citecolor=blue,          % color of links to bibliography
    filecolor=magenta,       % color of file links
    urlcolor=black           % color of external links
}

\usepackage[normalem]{ulem}

\def\sec#1{Sec.\ \ref{#1}}
\def\eq#1{Eq.\ (\ref{#1})}
\def\fig#1{Fig.\ \ref{#1}}

\journal{}

\begin{document}

\begin{frontmatter}

%-----------------------------------------------------------------

\title{Controlling thermoelectric, heat, and energy currents through a quantum dot in sequential and cotunneling Coulomb-blockade regimes} 
	
\author[a1]{Taha Yasin Ahmed}
\address[a1]{Division of Computational Nanoscience, Physics Department, College of Science, 
	University of Sulaimani, Sulaimani 46001, Iraq}

\author[a1,a2]{Nzar Rauf Abdullah}
\ead{nzar.r.abdullah@gmail.com}
\address[a2]{Computer Engineering Department, College of Engineering, Komar University of Science and Technology, Sulaimani, Iraq}

\author[a3]{Vidar Gudmundsson}
\ead{vidar@hi.is}
\address[a3]{Science Institute, University of Iceland, Dunhaga 3, IS-107 Reykjavik, Iceland}

%----------------------------------------------------------------

\begin{abstract}
Thermal transport through a Coulomb-blockade quantum dot (QD) coupled to two metallic leads is studied using five different approaches to the master equation in which sequential and coutuneling terms are taken into account. In the presence of intradot Coulomb interactions, a plateau in the thermo-particle, the heat, and the energy currents is seen. The current plateau diminishes at a high thermal bias between the leads. It is shown that the Pauli, the Redfield, the Lindblad-type
equation with first order tunneling rates, and first-order von-Neumann master equations give very similar thermal transport indicating the conservation of coherency in the electron transport in sequential tunneling between the QD and leads. In contrast, the thermal transport is suppressed when coutuneling processes are taken into account via a second-order von-Neumann master equation. The consideration of second order effects with respect to the QD-leads coupling brings
in a wealth of virtual processes at the contact to each lead.
These virtual processes directly weaken the effects of the
contribution of the first order direct processes to the overall transport, and introduce important other aspects of the transport, as level broadening, energy shifts, and lifetimes
in the time-domain. As a result the current plateau formed via the Coulomb interaction diminishes, when second order and cotunneling processes are considered.

\end{abstract}


\end{frontmatter}

\section{Introduction}

The thermoelectric (TE), properties of nanoscale device such as a quantum dot
have been attracting an increasing interest of research groups, as the QD devices 
will be useful in building up efficient energy conversion devices \cite{QD_Alivisatos}. 
The electrons or holes transfer through a material which can carry both charge
and energy. The TE phenomena thus include the transport of charge and energy, in the
form of heat. In bulk materials, TE parameters are not entirely independent
from each other. As a results, improvements using bulk TE material are limited \cite{Snyder2008}.
But the nanoscale systems can bring new regimes of length scales,
which permits the TE parameters to be tuned quasi-independently \cite{Heremans2013}. 
Nanoscale systems thus allow for an improvement of their TE properties. Dresselhaus realized that dimensional reduction inherent in a Bi quantum wire can improve their TE properties compared to bulk Bi \cite{PhysRevB.47.16631, PhysRevB.47.12727}.

There are many interesting phenomena in nanoscale QDs including the Coulomb blockade \cite{Wharam1995, Brotons-Gisbert2019}, and the Kondo effects \cite{LeHur2015}.
The charge quantization in a QD can lead to the important phenomenon called the Coulomb blockade \cite{TKuo2012}. The finger print of the Coulomb blockade on the TE properties has been studied \cite{doi:10.1063/1.4922907, JPC.25.465302}, and it has been shown that the Coulomb blockade leads 
to nonthermal broadening of the tunneling current peaks, a current rectification, coherent tunneling, and negative differential conductance in the Coulomb blockade regime \cite{ono2002current}. The inter-channel correlations in the Coulomb blockade that give rise to heat current reduction stem from the local charge conservation \cite{Sivre2018}.

Different versions of master equation have been used to study electron transport in Coulomb blockade QD coupled to electron reservoirs with Markovian and non-Markovian approaches \cite{e21080731, Vidar:ANDP201500298, doi:10.1021/acsphotonics.5b00532}. The master equation can be of different types such as the Pauli \cite{PAU28}, the Lindblad \cite{Lindblad1976}, the Redfield \cite{5392713}, and first/second-order von Neumann equations \cite{PEDERSEN2010595}. In these approaches it is possible to describe sequential and cotunneling electron transport between the QD and the electron reservoirs \cite{doi:10.1063/1.2927379}. The observation of cotunneling in QD system has recently been confirmed experimentally. It was shown that the cotunneling in a Coulomb-coupled double QD
is essential to obtain a correct qualitative understanding of the Coulomb drag \cite{PhysRevLett.117.066602}.

Spurn by the aforementioned studies, we use five different master-equation approaches for
the calculation of the nonlinear thermo-particle, the heat, and the energy currents through an interacting QD Coulomb blockade system, that takes into account the above-mentioned factors such as sequential and cotunneling mechanisms. 
We compare the thermal currents obtained via these master equations, and evaluate the conditions for all these types of master equation to study the thermal transport properties of the system.
In addition, we show that the Coulomb interaction in the QD leads to plateaus in all three types of mentioned thermal currents, when both the sequential and cotunneling processes are considered. 
It is shown that the current plateaus diminish in a slightly higher thermal energy bias between the leads.

The work is organized as follows: in \sec{section_model} the Hamiltonian of the total system, QD and leads, and the master equation formalism are presented. In \sec{section_results} the results of thermal transport under Coulomb interaction are displayed. In \sec{section_conclusion}, the main conclusion and remarks are presented.

\section{Model Hamiltonian and Time-evolution}\label{section_model}

In this section, the model and the main lines of the formalism used are presented.

\subsection{Model} 

Our model is a quantum dot attached to two metallic leads from the left and the right sides. Such a quantum dot device can be modeled by the following Hamiltonian \cite{PhysicaE.64.254}
\begin{equation}
	H_S = H_{\rm QD} + H_{l} + H_{\rm T},
	\label{equation_Hs}
\end{equation}
where $H_{\rm QD}$ is the Hamiltonian of quantum dot, $H_{l}$ is the Hamiltonian
for the external leads, and $H_{\rm T}$ describes the tunneling between the leads and quantum dot.
The first part of \eq{equation_Hs} is the Hamiltonian of the QD including the Coulomb 
interaction for the electrons, which can be defined as 
\begin{equation}
	H_{\rm QD} = \sum_i E_i \, d_i^{\dagger} d_i 
	           + \sum_{ijkl \atop i<j} U_{ijkl} \, d_i^{\dagger} d_j^{\dagger}  d_k d_l ,
\end{equation}
with $i < j$. Herein, $E_i$ indicates the energy of the single-electron (SE) state, $i$. $d^{\dagger}_i(d_i)$ is the creation (annihilation) operator of an electron in state $i$, and 
$U_{ijkl}$ are the Coulomb matrix elements in the SE state basis. 
The Coulomb interaction in the leads is neglected and the Hamiltonian of the leads shown in  \eq{equation_Hs} is
\begin{equation}
	H_{l} = \sum_{\alpha q} \varepsilon_{\alpha q} \, c^{\dagger}_{\alpha q} c_{\alpha q},
\end{equation}
where $\varepsilon_{\alpha q}$ are the single-particle energy spectra of the leads, and $c^{\dagger}_{\alpha q} (c_{\alpha q})$ creates(annihilates) an electron in the lead channel $\alpha$ with $q$ a continuous quantum number referring to momentum $\hbar q$.

The tunneling Hamiltonian defines the particle tunneling between the leads and the QD 
\begin{equation}
	H_{\rm T} = \sum_{\alpha q, i} t_{\alpha q,i} \, d_i^{\dagger} c_{\alpha q} + {\rm H.c.},
	\label{eq_tunneling}
\end{equation} 
where $t_{\alpha q,i}$ introduces the tunneling amplitude between the leads and the dot, and H.c. indicates the Hermitian conjugate of the first term of \eq{eq_tunneling}.
A significant energy scale in the calculations is the tunneling rate introduced as 
\cite{KIRSANSKAS2017317}
\begin{equation}
	\Gamma_{\alpha q, i}(E) = 2\pi \sum_{q} |t_{\alpha q, i}|^2 \delta(E-\varepsilon_{\alpha q}).
	\label{gamma_1}
\end{equation}
We note that all the physical calculated characteristics are scaled with $\Gamma$ as will become clear in subsequent sections.

\subsection{Time-evolution} 

In the calculation we employ units such that $\hbar = 1$, $k_B = 1$, and $|e| = 1$. Under this assumption, the electrical current of the system becomes a particle current. 
The time evolution of particles in the QD-leads system is calculated using a 
quantum master equation \cite{Haake1973,Breuer2002}. 
The reduced density operator describing the particles in the QD under the influence of the leads can be obtained from the total density operator of the QD-leads
by tracing out the variables of the leads \cite{Nzar_2016_JPCM, ABDULLAH2018}
\begin{equation}
	\rho_\mathrm{S}(t) = Tr_\mathrm{leads} \left\{ \rho (t) \right\},
\end{equation}
where $\rho$ the density operator of the full system. The QmeQ package is used to solve the master equation describing the time-evolution of reduced density operator numerically in the
interacting many-body Fock basis of the QD. Once the reduced density matrix is found, one can calculate the transport properties of the system as averages of the relevant operators
with the reduced density operator $\rho_\mathrm{S}(t)$. 
In QmeQ, two assumptions are taken into account \cite{KIRSANSKAS2017317}: First, the leads are considered thermalized according to the Fermi-Dirac occupation function $f_{\alpha}(E) = 1/[\exp{(E-\mu_{\alpha})/T_{\alpha}} + 1]$, where $T_{\alpha}$ and $\mu_{\alpha}$ are the temperature and the chemical potential of the leads. Second, both the left and the right leads have a constant density of states, $\nu(E) \approx \nu(E_F) = \nu_F$ with $F$ indicating the Fermi level. Under these assumption, the $q$-sums in \eq{gamma_1} are calculated as $\sum_{q} \rightarrow \nu_{F} \int^{+D}_{-D} dE$, where $D$ indicates the bandwidth of the leads. In addition, the tunneling amplitudes are energy independent, $t_{\alpha q, i} \approx t_{\alpha i}$. Using these conditions, one can write the tunneling rates as $\Gamma_{\alpha i} = 2\pi \, \nu_F \, |t_{\alpha i}|^2$.
An alternative approach retaining the energy dependence of the amplitudes and calculating 
the density of states in a quasi-one-dimensional leads in external magnetic field has been
taken elsewhere \cite{NewJournalOfPhysics.11.073019,Vidar11.113007}.

We are interested in studying the thermal properties of a QD in the steady state. So, we apply a thermal gradient to the QD via a temperature difference between the leads.
The thermo-particle current, TPC (I$^{\rm TPC}$), through the lead channel $\alpha$ can be defined as
\begin{equation}
	I^{\rm TPC}_{\alpha} =  -\frac{\partial}{\partial t} \expval{N_{\alpha}}
	= -i \expval{[H_{\rm S}, N_{\alpha}]},
\end{equation}
with $N_{\alpha} = \sum_q c^{\dagger}_{\alpha q} c_{\alpha q}$. 
The energy current, EC ($I^{\rm EC}_{\alpha}$), can also be found using
\begin{equation}
	I^{\rm EC}_{\alpha}  = -\frac{\partial}{\partial t} \expval{H_\alpha}
	= -i\expval{[H_{\rm S}, H_{\alpha}]},
\end{equation}
where $H_{\alpha} = \sum_q c^{\dagger}_{\alpha q} c_{\alpha q}$.
Finally, the heat current, HC ($I^{\rm HC}_{\alpha}$), emanating from the lead channel $\alpha$ is defined as
\begin{equation}
 I^{\rm HC}_{\alpha} = I^{\rm EC}_{\alpha} - \mu \, I^{\rm TPC}_{\alpha},
\end{equation}
where $\mu$ is the chemical potential of the leads.

\section{Results}\label{section_results}

\begin{figure}[htb]
	\centering
	\includegraphics[width=0.35\textwidth]{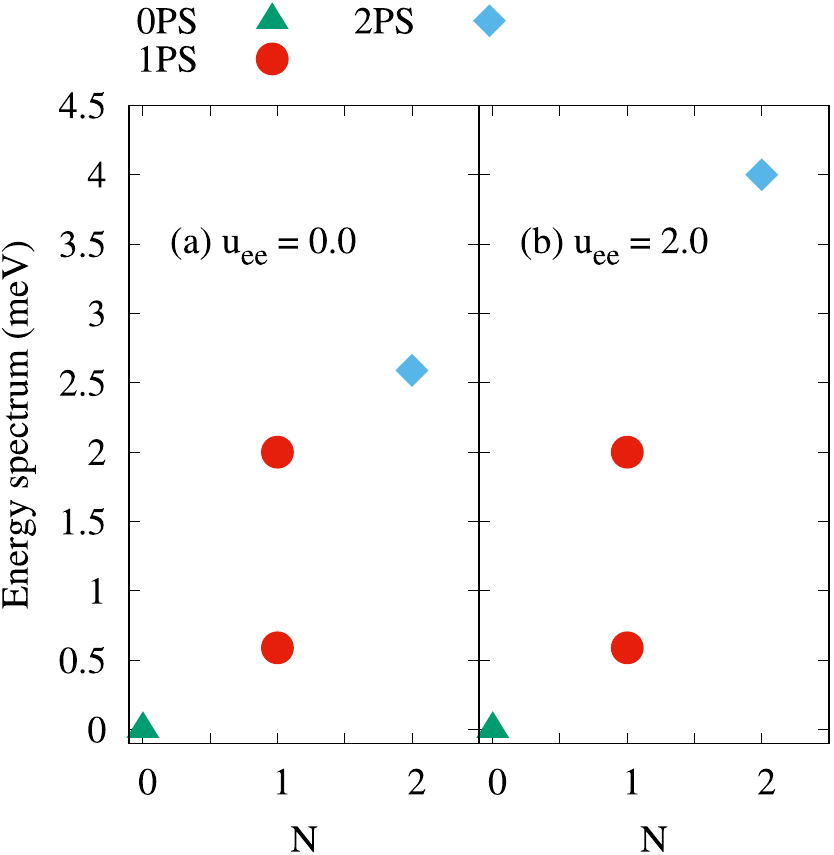}
	\caption{Energy spectrum (Many-body energy spectrum) as a function of particle number, $N$, without, $u_{ee} = 0.0$ meV, and with, $u_{ee} = 2.0$ meV, Coulomb interaction, where $u_{ee}$ is strength of the Coulomb interaction. 
	A triangle (green) indicates zero-particle state, 0PS, Circles (red) the one-particle states, 1PS, and Diamond (blue) the two-particle state, 2PS.}
	\label{fig01}
\end{figure}

Our results for the TPC, HC, and EC are presented in this section tuning the thermal energy of the leads, the Coulomb interaction between electrons in the QD, and applying different approaches to the master equation describing the time-evolution. The many-particle states for the QD without (a), and with (b) the Coulomb interaction are displayed in \fig{fig01}. 

In the absence of the Coulomb interaction, $u_{ee} = 0.0$, the QD has one zero-particle state, 0PS, with energy $E_0^0 = 0.0$ (green triangle), two one-particle states with energy values of $E_1^0 = 0.59$ and $E_1^1 = 2.0$~meV, and one two-particle state located at $E^0_2 = 2.59$~meV. Herein, the subscript shows particle number and the superscript indicates the state (ground-state, $0$, or excited-state, $1$) of each of the particles states.   

In the presence of the Coulomb interaction with strength $u_{ee}=2.0$ meV, the zero- and the one-particle states remain unchanged while the two-particle state is shifted up to $\check{E}_2^0 = E^0_2 + \sqrt{u_{ee}} = (2.59 + \sqrt{2})\: \mbox{meV} = 4.0$~meV. 

In order to generate thermal transport via the QD, we apply a temperature gradient between the leads, where the thermal energy of the left(right) lead is assumed to be $0.25$($0.1$)~meV. First, we use the first-order von-Neumann master equation (1vN) \cite{Goldhaber-Gordon1998}. 
The evolution of the whole system (leads and QD) is described by the full density operator, $\rho$, via \cite{GUDMUNDSSON20181672}
\begin{equation}
	\frac{\partial \rho}{\partial t} = -i [H_{\rm S}, \rho],
\end{equation}
where the density matrix elements for the sectors in the Fock space determined by the number of particle (electron/hole) excitation in the leads are grouped together \cite{Goldozian2016}. In the 1vN method, all the density matrix elements with more than two particle excitations are neglected. In addition, it is also assumed that the particles in the leads are thermally distributed according to the Fermi–Dirac equilibrium distribution, $f_{\alpha}(E)$, and this distribution is not affected by the coupling to the quantum dot. The leads play the role of the external reservoirs. 
\begin{figure}[htb]
	\centering
	\includegraphics[width=0.4\textwidth]{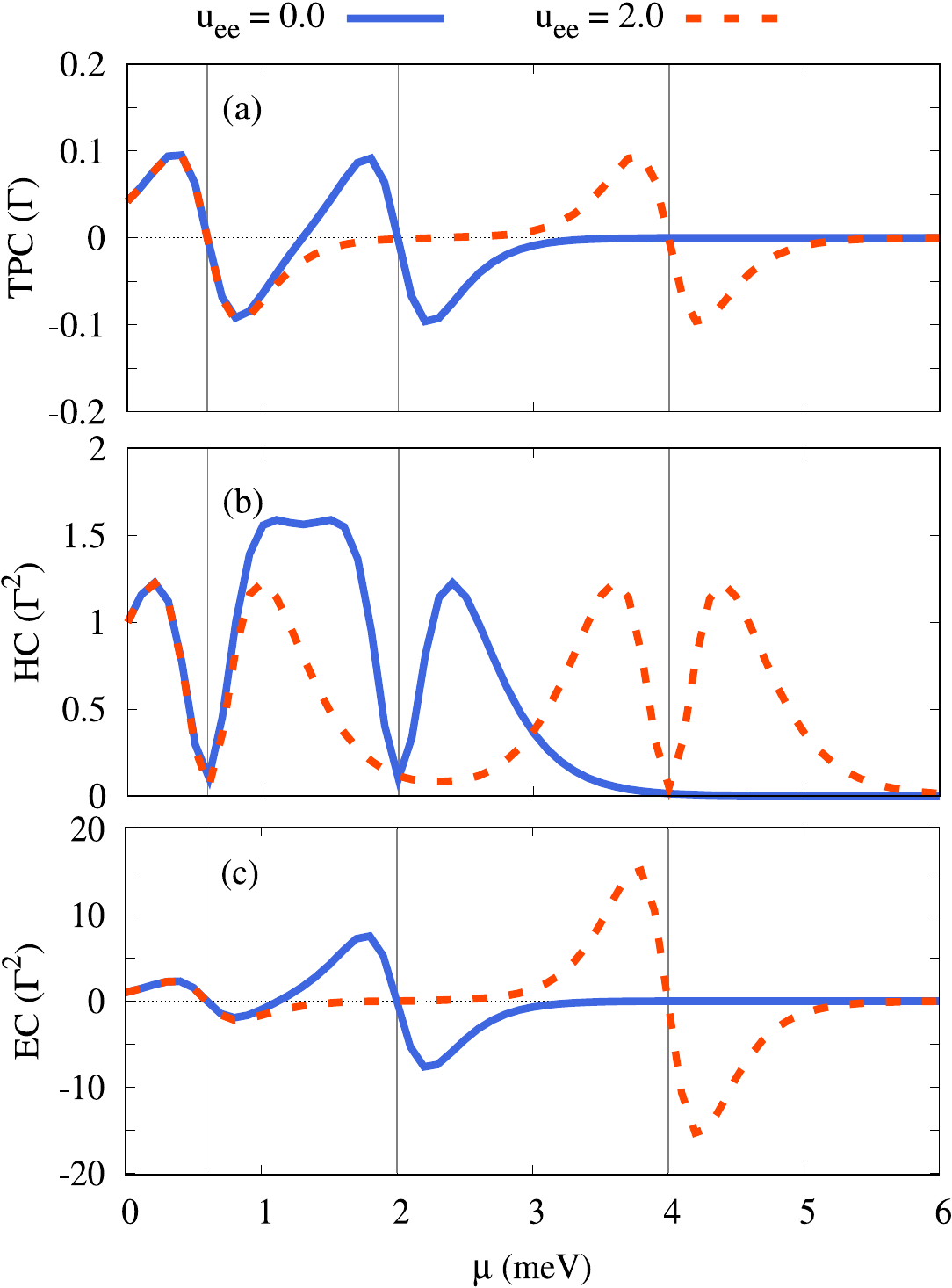}
	\caption{(a) Thermo-particle current, TPC, (b) heat current, HC, and (c) energy current, EC, versus chemical potential of the leads, $\mu$, without (blue) and with (red) the Coulomb interaction with strength of $u_{ee} = 0.0$, and $2.0$~meV, respectively. The thermal energy of the left lead and the right lead are assumed to be $k_B T_{L} = 0.25$ meV, and $k_B T_R = 0.1$ meV, respectively. The strength of the tunneling rate is $\Gamma_L = \Gamma_R = 0.0025$~meV, and the 1vN is used.}
	\label{fig02}
\end{figure}
In summary, the below approximations are applied in the 1vN approach: First, terms involving up to a single excitation are allowed. Second, a Markov approximation is considered. Using these assumption, one can realize the following properties of the 1vN master equation: First, the 1vN includes only sequential tunneling in the presence of coherences. Second, the coupling strength between leads and the QD has to be smaller than the temperature of the leads $\Gamma_{L,R} << T_{L,R}$. Third, the 1vN can violate the positivity of $\rho_\mathrm{S}$, the reduced density operator \cite{Goldhaber-Gordon1998}. 
The above description can be rephrased in the projection formalism for the construction of
the master equations by indicating to which order the memory kernel of the master equation is expanded
with respect to the coupling between the leads and the central system, the QD \cite{Haake1973,Breuer2002}.

In \fig{fig02}, the thermo-particle current (TPC) (a), the heat current (b), and the energy current (c) versus the common chemical potential of the leads are calculated for the QD without (blue solid line) and with the Coulomb interaction (red dashed line) of strength $u_{ee} = 2.0$ meV. 
The black vertical lines represent the position of states shown in \fig{fig01}.  
We use the 1vN master equation with the condition that $\Gamma_{L,R} << T_{L,R}$. The temperature of the leads is considered to be $T_L(T_R) = 0.25$($0.1$), and $\Gamma_{L,R} = 0.0025$.

\begin{figure}[htb]
	\centering
	\includegraphics[width=0.4\textwidth]{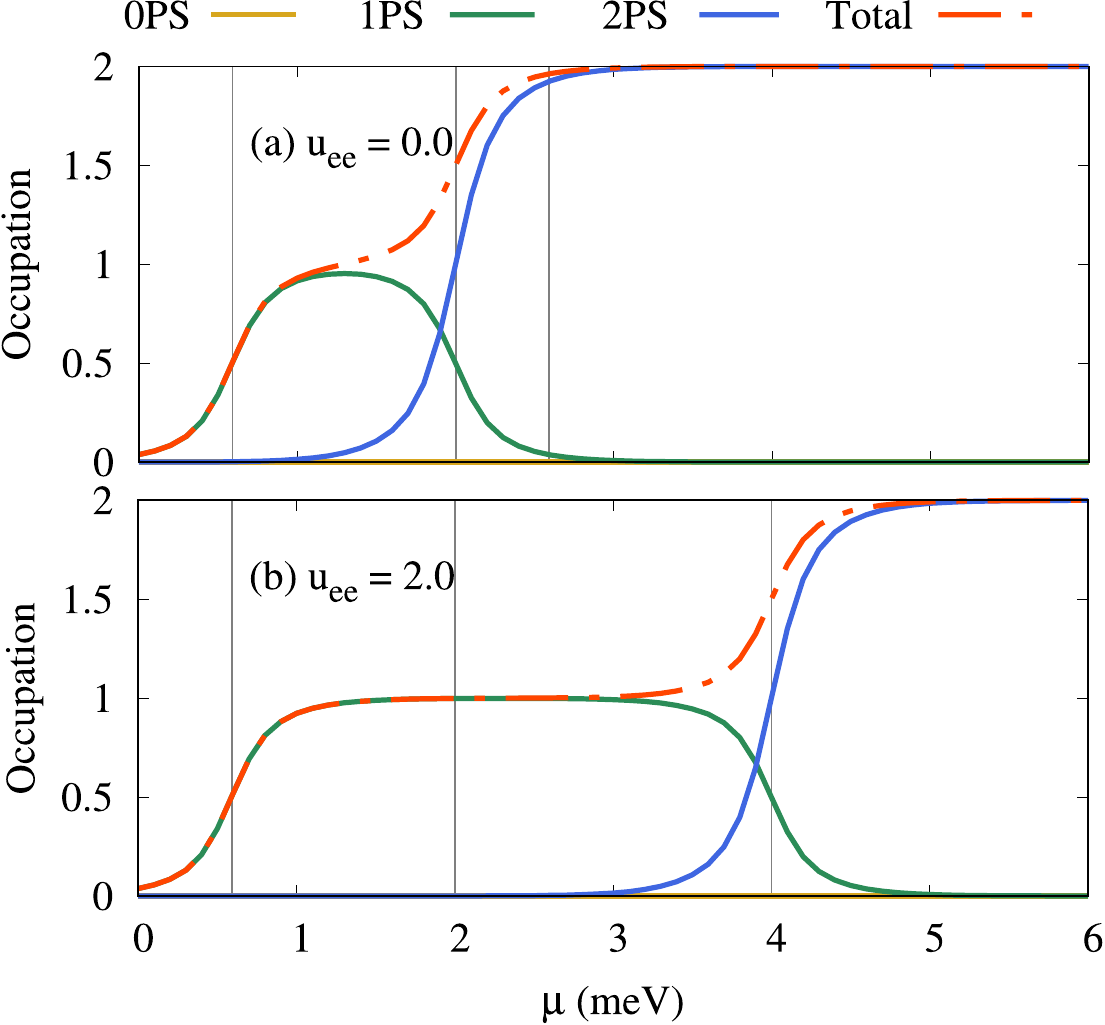}
	\caption{Occupation as a function of the chemical potential without (a) and with (b) the Coulomb interaction of strength $u_{ee} = 0.0$, and $2.0$~meV, respectively. Golden solid line indicates charging for the zero-particle states, green solid line is the charging for the one-particle states, blue solid line displays the charging for the two-particle states, and red dashed dot line is the total charging of the system. The thermal energy of the left lead and the right lead are assumed to be $k_B T_{L} = 0.25$ meV, and $k_B T_R = 0.1$ meV, respectively.  The strength of tunneling rate is $\Gamma_L = \Gamma_R = 0.0025$~meV, and the 1vN is used.}
	\label{fig03}
\end{figure}

In the absence of the Coulomb interaction in the QD, the TPC, HC, EC are zero at $\mu = 0.59$, and $2.0$~meV, which are the locations of the one-particle states shown in \fig{fig01}(a). These locations correspond to half filling of states. We note that between these states at $\mu = 1.29$~meV, the TPC and EC are again zero when an integer filling is attained. In order to show the half and integer filling of the states, we present the occupation of the QD without (a) and with (b) the Coulomb interaction in \fig{fig03} for all states in the calculation.

In \fig{fig03}(a), the occupation or the charging of the zero-, one-, and two-particle states is presented for a QD with no Coulomb interaction. At the chemical potential $\mu = 0.59$~meV the 1PSs is half charged, and at $\mu = 2.0$~meV, the 2PS and 1PSs are half charged and discharged, respectively. At the point of $\mu = 2.59$~meV, which is the location of the 2PS, the QD is almost charged 
by an integer of particles as the Coulomb interaction is zero \cite{TAGANI201336_PAT}.

In the presence of the Coulomb interaction, $u_{ee} = 2.0$ meV, broad plateaus of zero TPC, HC, and EC around $\mu = 2.0$~meV are found (red dotted line). This location corresponds to an integer filling of the 1PSs (see \fig{fig03}(b)), and the total charge (red dash dotted line) indicates an integer filling. In addition, the 2PS becomes active in the case of the Coulomb interaction at $\mu = 4.0$~meV in which a high TPC, HC, and HC around the 2PS and zero value of TPC, HC, and HC at the $\mu = 4.0$~meV are seen.
The activation of the 2PS and the current plateaus are caused by the Coulomb interaction. 
This can be confirmed by the occupation of the 2PS shown in \fig{fig03}(b) in which a half filling of 
the 2PS (blue line) occurs at $\mu = 4.0$~meV, and at the same time a discharging of the 1PSs is seen. 
We have earlier reported the effects of the Coulomb interaction on the TPC in a QD and its current plateaus \cite{PhysicaE.53.178}.

\begin{figure}[htb]
	\centering
	\includegraphics[width=0.45\textwidth]{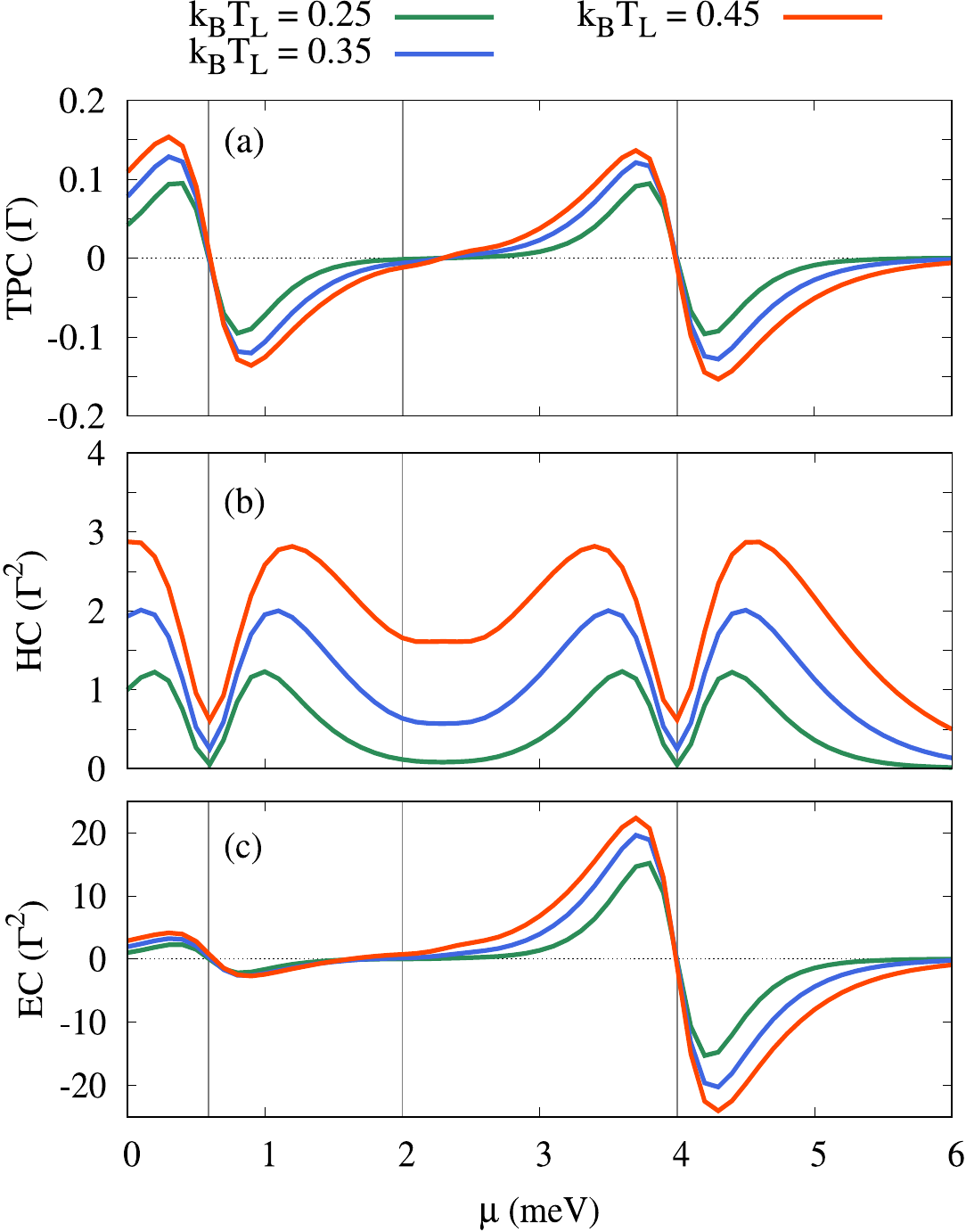}
	\caption{Thermoelectric current, TEC, (b) heat current, HC, and (c) energy current, EC, versus chemical potential of the leads, $\mu$ in the presence of a Coulomb interaction with strength of $2.0$~meV. 
		The thermal energy of the right lead is fixed at  $k_B T_R = 0.1$~meV and the thermal energy of the left lead is assumed to be $k_B T_{L} = 0.25$ (green), $0.35$ (blue), and $0.45$~meV (red). The strength of tunneling rate is $\Gamma_L = \Gamma_R = 0.0025$~meV,
		and a 1vN is used.}
	\label{fig04}
\end{figure}

The thermal currents, PTC (a), HC (b), and EC (c), for different temperature of the leads are presented in \fig{fig04}, where the Coulomb interaction in the QD is considered. The thermal energy of the right lead is fixed at $0.1$~meV, and the thermal energy of the left lead is changed to $0.25$ (green), $0.35$ (blue), and $0.45$~meV (red).

All three types of thermal currents through the QD increase with increasing temperature gradient or thermal energy of the leads. The zero value of the HC corresponding to the half filling located at the energy of the 1PS and 2PS states is affected and increases at higher thermal energy of the leads. It is also interesting to see that the current plateaus almost vanish at the higher thermal energy, $k_{\rm B}T_{\rm L} = 0.45$~meV, at $\mu = 2.0$ meV.

\begin{figure}[htb]
	\centering
	\includegraphics[width=0.45\textwidth]{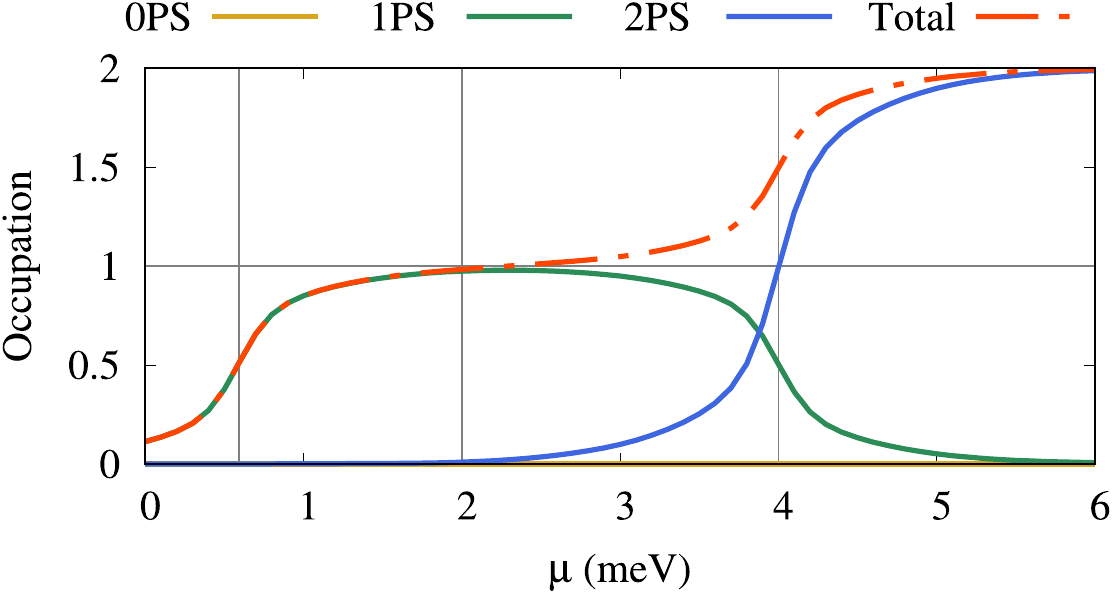}
	\caption{Occupation as a function of the chemical potential with a Coulomb interaction, where the thermal energy of the leads is $k_{\rm B}T_{\rm L} = 0.45$~meV, and $k_B T_{\rm R} = 0.1$ meV.   Golden solid line indicates charging of the 0PS, green solid line is the charging for the 1PSs, blue solid line displays the charging for 2PS, and red dashed dot line is the total charging of the system.  The strength of tunneling rate is $\Gamma_L = \Gamma_R = 0.0025$~meV, and the 1vN is used.}
	\label{fig05}
\end{figure}

The properties of the thermal currents can be explained using the occupation of the QD, which is shown 
in \fig{fig05} where the thermal energy of the left lead is $0.45$~meV and the right lead is $0.1$~meV. We notice a thermal smearing, as the occupation of the 1PS considerably deviates from 1, while the occupation of the 2PS increases considerably from zero around $\mu = 2.0$~meV.
Consequently, the 1PSs contribute a negative current, whereas the 2PS support a positive current resulting in a positive total current.

It is necessary to contrast the thermal currents to the current generated by a bias voltage instead of temperature gradient of the leads. In this case the thermal energies of the leads are fixed at $0.1$~meV, but the electrical bias voltage is generated via $V_{\rm bias} =  \mu_L - \mu_R = 0.15$~meV.
\begin{figure}[htb]
	\centering
	\includegraphics[width=0.45\textwidth]{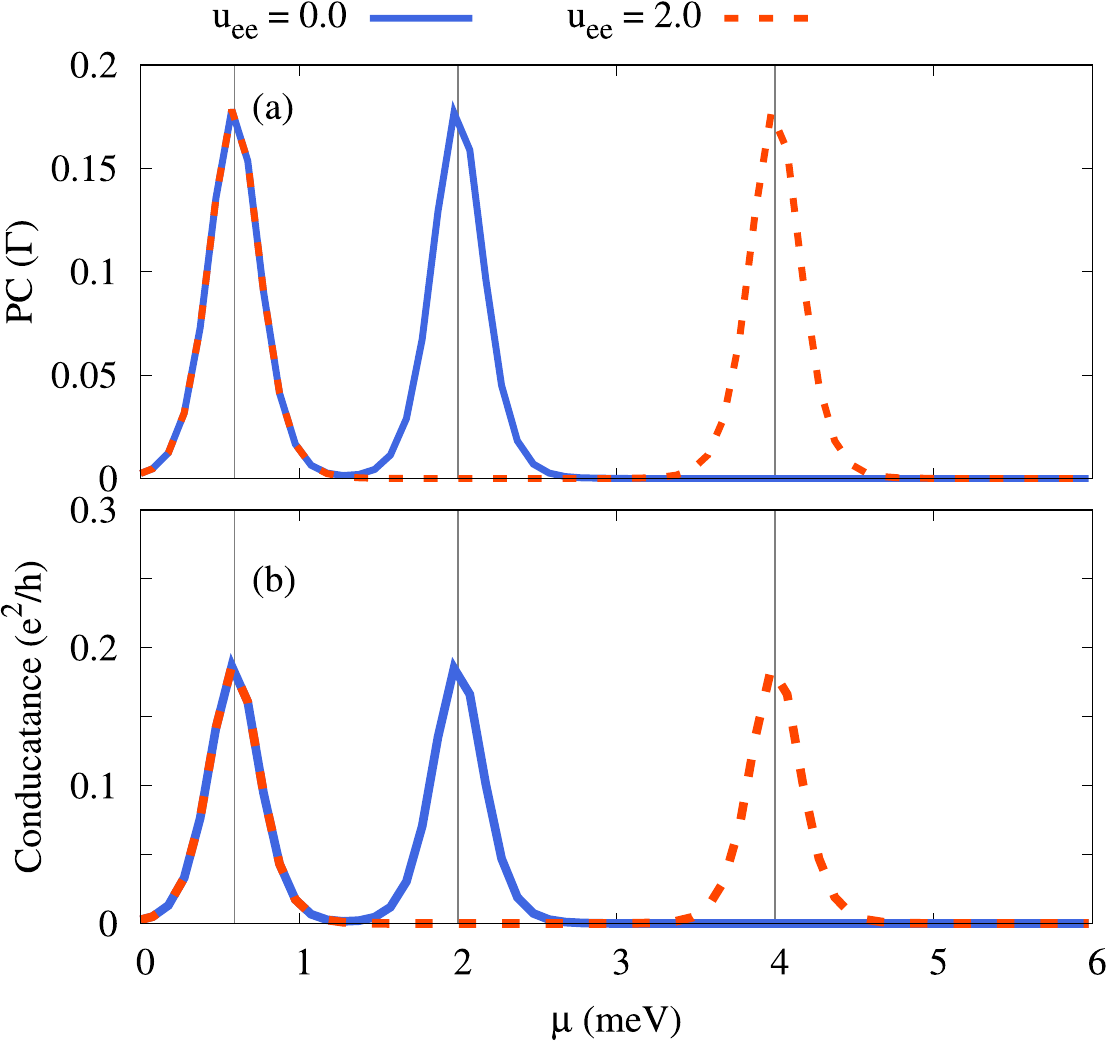}
	\caption{(a) Particle current and (b) conductance as a function of chemical potential, $\mu$,  under the electrical bias voltage $\mu = \mu_L - \mu_R = 0.15$~meV for the system without (blue) and with (red) Coulomb interaction of strength $u_{ee} = 0.0$, and $2.0$~meV, respectively. The thermal energy of both leads is fixed at $k_B T_{L,R} = 0.1$~meV. The strength of tunneling rate is $\Gamma_L = \Gamma_R = 0.0025$~meV, and the 1vN is used.}
	\label{fig06}
\end{figure}
The particle current, PC (a), and the conductance (b) are shown in \fig{fig06} for the QD without (blue) and with (red) the Coulomb interaction. The chemical potentials of the leads are changed, but the difference between chemical potentials is kept fixed, $\mu_L - \mu_R = 0.15$~meV, in each steps. It is clear that the PC is positive, with peaks reflecting the half filling of the states, and with zeros between the peaks, an indication of the Coulomb blockade in the transport. The enhanced current plateaus in both the PC and the conductance are again seen at $\mu = 2.0$~meV caused by the Coulomb interaction. 

Next we study the different types of master equation approaches, such as the Pauli \cite{Grabert1982}, the Redfield \cite{PhysRev.89.728}, the Lindblad equation with first order tunneling rates \cite{PEDERSEN2010595}, the 1vN, and the second-order von Neumann (2vN) \cite{PhysRevB.72.195330} for investigating thermal transport through the QD. We prefer to compare the thermal currents obtained via these different types of master equations. 
The results for the 1vN are already shown fulfilling the condition of weak coupling between the QD and the leads in the Markovian approximation. The Markovian approximation is justified by our 
interest in the steady state \cite{JONSSON201781}. 
In the weak coupling regime the Pauli, Redfield, and Lindblad equation with first order tunneling rates \cite{Lindblad1976} approaches lead to similar results as the 1vN. In these approaches only single particle excitation is taken into account. Second: The positivity of reduced density matrix is implicitly preserved. Third: The coupling strength is weak and $\Gamma_{L,R} << T_{L,R}$.  
Sequential tunneling in the presence of coherences is taken into account for the Redfield and the Lindblad approaches, while the coherences in the Pauli formulation are neglected \cite{breuer2002theory}.
The Markov approximation is used for Pauli and the Lindblad approach, while the Markov approximation in the Redfield formulation is accompanied with further conditions listed in the documentation of the QmeQ-package \cite{KIRSANSKAS2017317}.

\begin{figure}[htb]
	\centering
	\includegraphics[width=0.45\textwidth]{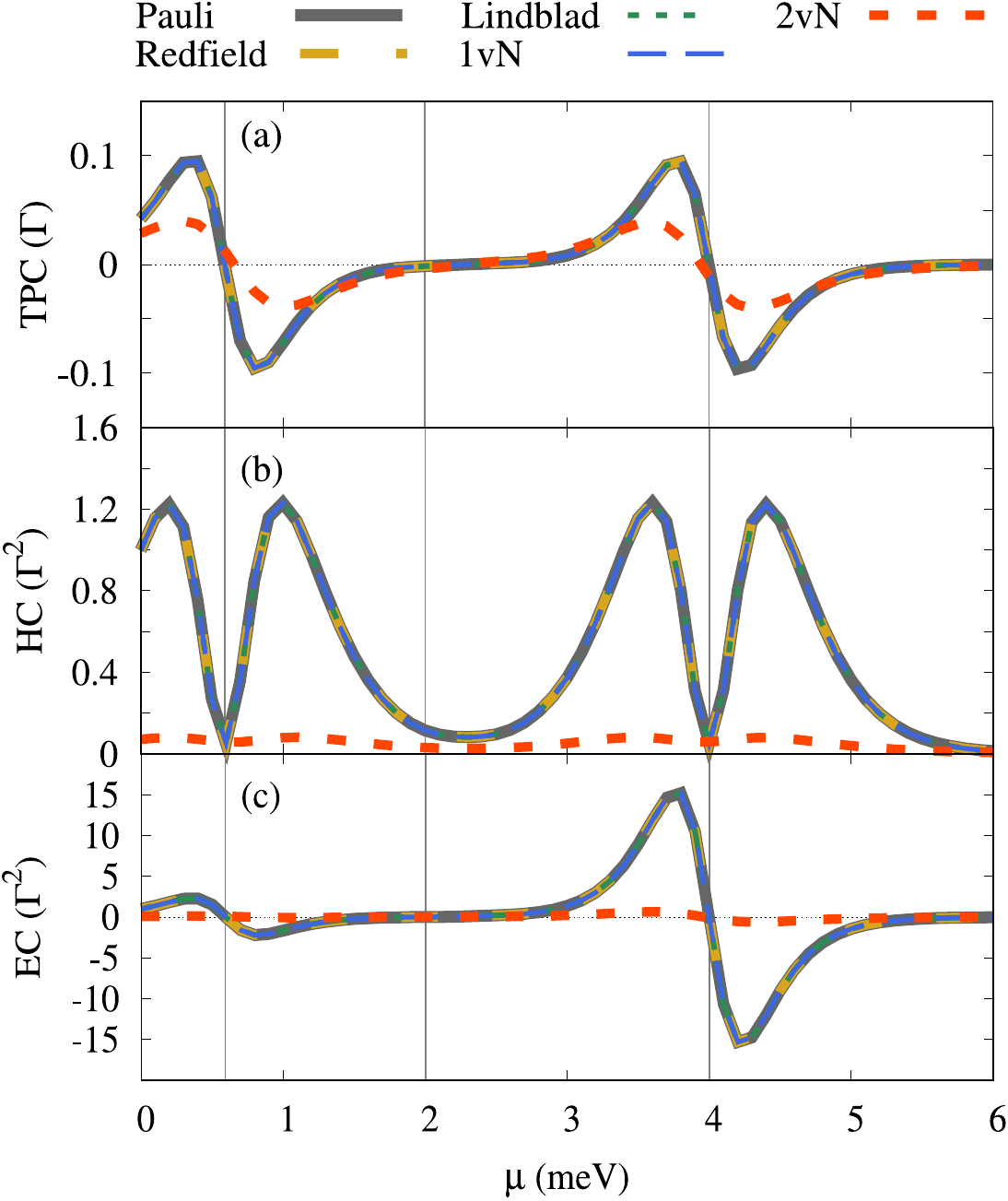}
	\caption{(a) Thermoelectric current, TEC, (b) heat current, HC, and (c) energy current, EC, versus chemical potential of the leads, $\mu$, in the presence of Coulomb interaction with strength of $u_{ee} = 2.0$.  The  Pauli (gray line), Redfield (golden dotted dash), Lindblad (dotted green), 1vN (dashed blue), 2vN (dashed red) master equations are used.
		The thermal energy of the left lead and the right lead are assumed to be $k_B T_{L} = 0.25$ meV, and $k_B T_R = 0.1$ meV, respectively. The strength of tunneling rate for Pauli, Redfield, Lindblad, and 1vN is  $\Gamma_{L,R} = 0.0025$~meV, and $0.25$~meV is used for 2vN method.}
	\label{fig07}
\end{figure}

The 2vN approach for the  master equation is different from the four aforementioned methods. 
In the 2vN method, 
the approximations are the same as for the 1vN method except the terms including up to two particle excitations are taken into account, and the condition of the coupling strength is $\Gamma_{L,R} \lesssim T_{L,R}$ \cite{PEDERSEN2010595}. Conseqently, the 2vN can include sequential tunneling, cotunneling, pair-tunneling, and broadening effects \cite{PhysRevB.72.195330}.

In \fig{fig07}, the TPC, HC, and EC versus the chemical potential of the leads in the presence of the Coulomb interaction in the QD are plotted for the Pauli (gray line), Redfield (golden dotted dash), Lindblad (dotted green), 1vN (dashed blue), 2vN (dashed red) master equations. In order to satisfy the necessary condition for the Pauli, Redfield, and Lindblad equations, we consider $\Gamma_{L,R} = 0.0025$~meV, much smaller than the temperature of the leads, $T_L$. Also, we assume the coupling strength of the leads is $\Gamma_{L,R} = 0.25$, which satisfies the condition for the 2vN master equation, $\Gamma_{L} \approx T_L$.  

One can clearly see that the TPC, HC, and EC are very similar for the Pauli, Redfield, Lindblad, and the 1vN master equations, which indicates that the transport coherency is kept in our system under the applied selected range of temperature gradient between the leads. There is even no deviation for these equations for the TPC, HC, and the EC at the location of the 1PS and the 2PS (black vertical lines). 

It is interesting to see the results of the 2vN approach in \fig{fig07} in which the amplitudes for the TPC, HC and the EC are decreased. 
\begin{figure}[htb]
	\centering
	\includegraphics[width=0.45\textwidth]{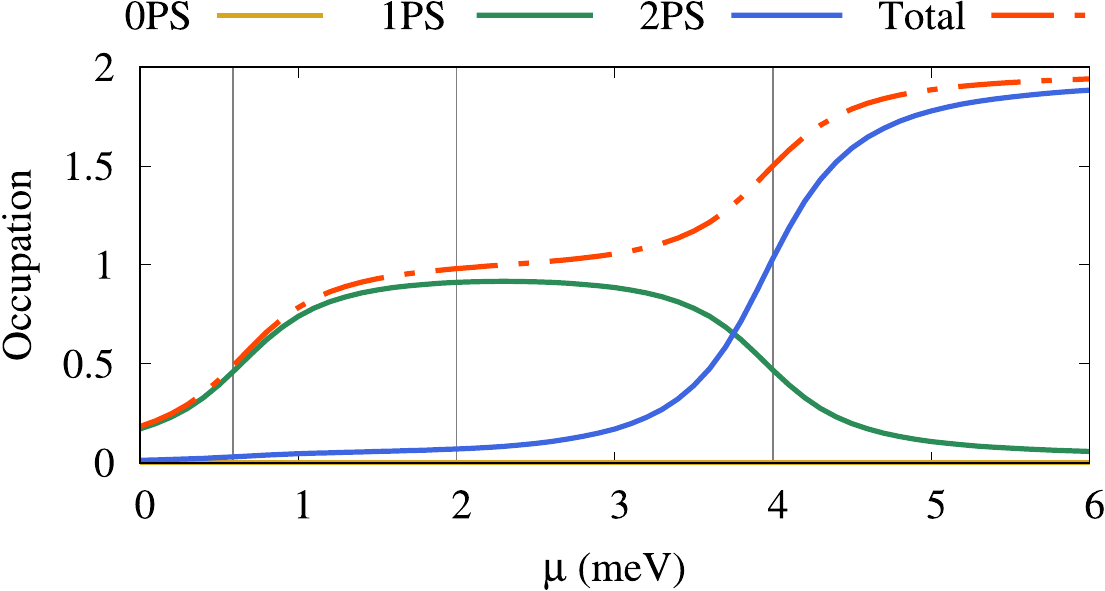}
	\caption{Occupation of the QD states as a function of the chemical potential in the presence of the Coulomb interaction of strength $u_{ee} = 2.0$ meV for the 2vN master equation with $\Gamma_{L,R} = 0.25$~meV. The golden solid line indicates charging for the zero-particle state, the green solid line is the charging for the one-particle states, blue solid line displays the charging for the two-particle states, and red dashed dot line is the total charging of the system. The thermal energy of the left and the right leads are assumed to be $k_B T_{L} = 0.25$ meV, and $k_B T_R = 0.1$ meV, respectively.
	}
	\label{fig08}
\end{figure}
This softening of the ``thermodynamical response'' of the quantum dot described by the 2vN
master equation compared to the 1st order equations can clearly not be referred to a 
difference in the occupation of the states of the QD, as can be seen in \fig{fig08}.
Instead, the consideration of second order effects with respect to the QD-leads coupling
brings in a wealth of virtual processes at the contact to each lead.  
These virtual processes directly weaken the effects of the contribution of the 
first order direct processes to the overall transport, and introduce important
other aspects of the transport, as level broadening, energy shifts, and lifetimes 
in the time-domain.

\begin{figure}[htb]
	\centering
	\includegraphics[width=0.45\textwidth]{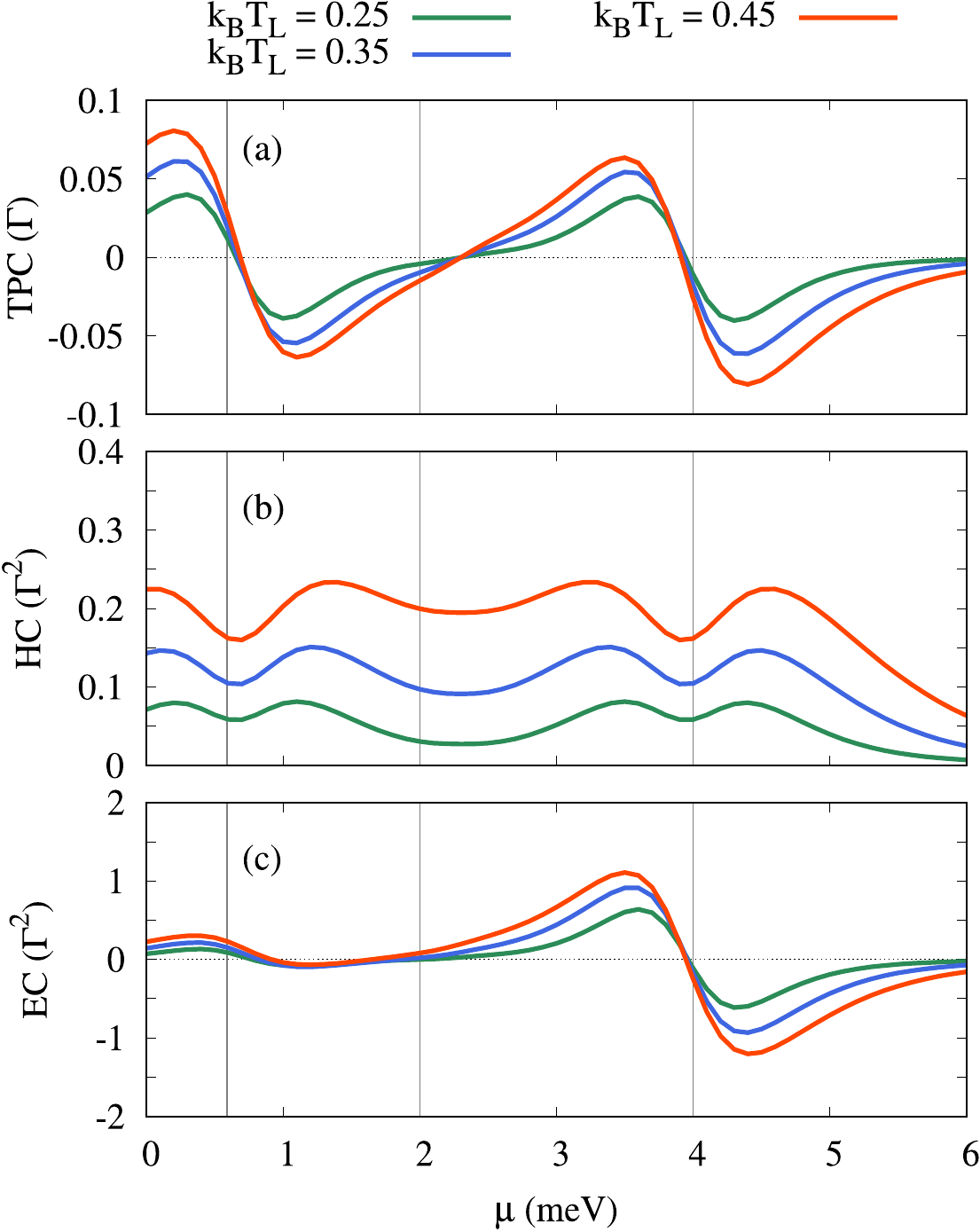}
	\caption{Thermoparticle current, TPC, (b) heat current, HC, and (c) energy current, EC, versus the chemical potential of the leads, $\mu$ in the presence of a Coulomb interaction with the strength of $2.0$~meV. The thermal energy of the right lead is fixed at  $k_B T_R = 0.1$~meV and the thermal energy of the left lead is assumed to be $k_B T_{L} = 0.25$ (green), $0.35$ (blue), and $0.45$~meV (red). The 2vN is used with $\Gamma_{L,R} = 0.25$~meV.}
	\label{fig09}
\end{figure}

Figure \ref{fig09} clearly shows the importance of the thermodynamic transport through
the QD for the 2vN case, and how it depends on the ``thermal bias'' put on the quantum dot.
A close inspection of \fig{fig09} gives an indication of the level broadening 
in the system through the softening of the main features and the shift of zeros and
minima away from the exact energies of the original levels of the closed QD.
So, by no surprise we realize here how important second order effects in the tunneling
process between the leads and the QD can be.

\section{Conclusions and Remarks}\label{section_conclusion}

In this work, we have studied thermal transport through a QD generated via a thermal bias between the two metallic leads connected to the QD using several types of master equations. The intradot Coulomb interaction in QD is taken into account, and it is seen that the influence of the intradot Coulomb interaction is not the same for both the sequential and the cotuneling processes considered for the QD-leads coupling.
The thermo-particle, the heat, and the energy currents are suppressed in the presence of 
second order and cotuneling processes. This can be related to the virtual processes at the contact region of the QD-leads. The virtual processes may include level broadening and energy shifts. It thus leads to less or slightly different charging of the states of the QD compared to the case where only sequential first order processes are included. Consequently, the suppression of the thermal current for the case of models with higher order QD-lead couplings and cotunneling is seen.

\section{Acknowledgment}

This work was financially supported by the University of Sulaimani and 
the Research center of Komar University of Science and Technology. 
The computations were performed on resources provided by the Division of Computational 
Nanoscience at the University of Sulaimani.  
 
%\section{References}

%\bibliographystyle{elsarticle-num} 
%\bibliography{Ref.bib}

\end{document}